\documentclass{elsart}

\usepackage{psfig}

\usepackage{natbib}

\begin{document}

\runauthor{Miller et al.}

\begin{frontmatter}

\title{Evidence of Rapid Optical Variability In Selected Narrow-Line
Seyfert 1 Galaxies}

\author{H.R.Miller, E.C.Ferrara, J.P.McFarland, 
J.W.Wilson, A.B.Daya}
\address{Department of Physics \& Astronomy, 
Georgia State University, Atlanta, GA 30303 USA}

\author{R.E. Fried}
\address{Braeside Observatory, P. O. Box 906
Flagstaff, AZ 86002 USA}

\begin{abstract}
We present the first results of a search for the presence of rapid
optical variability in a sample of five Narrow-Line Seyfert 1 galaxies.  We find
clear evidence of rapid variability for IRAS 13224-3809 with variations occurring on
time scales of an hour.  However, the results are less conclusive for the
other four sources
in our sample, Markarian 766, PG 1244+026, PG 1404+226 and Arakelian 564.  While there
are several instances among these latter objects where there is a hint that
variability 
may be present,
IRAS 13224-3809 provides the only conclusive evidence of rapid optical variability 
detected to date.

\end{abstract}


\end{frontmatter}


\section{Introduction}

Narrow-line Seyfert 1 galaxies (NLS1) are an unusual and unique class of AGN
which exhibit extreme X-ray properties.  These galaxies are characterized
by the following properties (Boller, Brandt and Fink 1996):  
(1) they have Balmer lines of 
hydrogen that are only slightly broader 
than the forbidden lines in their spectrum, (2) they tend to have strong Fe II and 
weak [O III] lines, and they thus
tend to lie at one extreme of the Boroson \& Green Eigenvector (Boroson and Green 1992),
(3) X-ray observations have shown the existence of persistent giant-amplitude variability
observed at soft X-ray energies, and (4) they exhibit a strong, soft X-ray excess 
emission and a steep 2-10 keV 
power-law continuum.  The X-ray observations 
indicate that Narrow-Line Seyfert 1s are not rare and represent a substantial part of the 
Seyfert population. The origin of the distinctive X-ray/optical properties is as yet
unknown, but it has been suggested that it is linked to an extreme value of a 
primary physical parameter; e.g.  
the accretion rate, nuclear orientation and black hole spin have been suggested.

One of the intriguing characteristics of NLS1 galaxies is
the existence of the persistent giant-amplitude variability
observed at soft X-ray energies.  Although variability is a well-known X-ray characteristic
of Seyfert 1 galaxies,  the extreme variability reported for 
IRAS 13224-3809 (Boller et al. 1997) is
astonishing.  As a result of these reports of extreme variability, 
we recognized that it would be important to determine if a similar phenomenon is present at
optical wavelengths, and if so, what is the relationship between 
the variability observed in these different regimes.

\section{Observations}

The high-time-resolution observations of the NLS1s 
reported here were
obtained with the 1.05-meter telescope at Lowell
Observatory, the 0.6-meter telescope at Siding Spring Observatory, the 0.4-meter 
telescope at Braeside Observatory, and the 0.4-meter telescope at GSU's Hard Labor 
Creek Observatory.  The
observations were made with CCD cameras at each observatory equipped with 
a $BVRI$-filter set.  Repeated exposures of 90-120 seconds were obtained for 
the star field containing 
the NLS1 galaxy and several comparison stars.  These comparison
stars were internally calibrated and are located on the same CCD
frame as the object of interest.  They were used as the reference standard stars in the data 
reduction process.  The observations were reduced
using the method of Howell and Jacoby (1986). Each exposure is
processed through an aperture photometry routine which reduces the data
as if it were produced by a multi-star photometer. Differential
magnitudes can then be computed for any pair of stars on the frame.
Thus, simultaneous observations of the NLS1, several comparison stars,
and the sky background will allow one to remove variations which may be
due to fluctuations in either atmospheric transparency or extinction.
The aperture photometry routine used for these observations is either the CCDPHOT
package
or the {\it
apphot} routine in IRAF.

The errors are determined from the scatter in the differential photometry of
two comparison stars observed simultaneously with the object of interest.  
The length of the vertical bars attached to each data point corresponds to a 
2$\sigma$ error bar.


\begin{figure}
\centerline{\psfig{figure=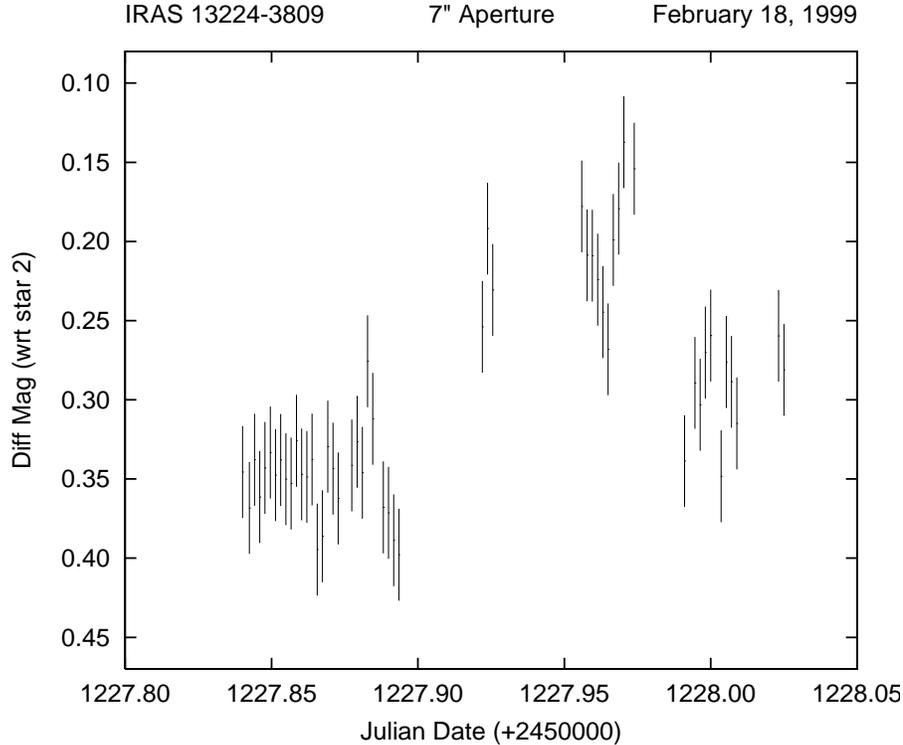,width=12cm,angle=0,clip=}}
\caption{The rapid optical variability of IRAS 13224-3809 is clearly present
on the night of 1999 Feb 18.  There is evidence for the presence of three flares with
amplitudes ranging from 0.1 - 0.25 mag. occurring during
the approximately 5 hours of monitoring on this night.}
\end{figure}


\section{Results}

During the past two years, we have embarked upon a program to
monitor the rapid optical variability 
of a sample of five
NLS1 galaxies.  The primary goal of this program is the detection of
rapid optical variations which may be the optical counterpart 
of the giant X-ray variability.  We have recently 
determined that optical variations
of smaller amplitude, but on similar timescales, to those seen at X-ray wavelengths 
are clearly present for
IRAS 13224-3809.  However, the results are less conclusive
for the other four galaxies in our sample (AKN 564, MRK 766,
PG 1404+226 and PG 1244+026).  

IRAS 13224-3809 was monitored for rapid optical variability 
for five nights in 1999 February.  The most extreme variability detected is displayed
in Fig. 1.  The object was monitored for approximately 5 hours on this 
night with the detection of three events:  event 1 occurred at 1227.88 JD with an
approximate amplitude of 0.12 mag; event 2 occurred at 1227.925 JD with an approximate
amplitude of 0.2 mag; and event 3 occurred at 1227.97 JD with an approximate
amplitude of 0.2 mag.  The time scales of these flares are comparable to the time
scales of the most rapid X-ray flares observed, but the 
amplitudes are substantially smaller.
These events may be the result of small discrete events,
such as hot spots or flares in the accretion disk, or may be due to magnetic reconnections
(Mineshige et al, 2000).

\section{Acknowledgements}

The authors wish to thank Mount Stromlo and Siding Spring Observatory and 
Lowell Observatory for their
generous allocations of observing time 
in support of this program.   HRM, ECF,ABD and JPM
are grateful for support from grants from the Research
Corporation, NASA grant NAGW-4397, and the PEGA-RPE program at Georgia State 
University.


\end{document}